\documentclass[10pt,aps,pra,showkeys,twocolumn,superscriptaddress]{revtex4-2}
\usepackage{amsmath, amssymb}
\usepackage{mathrsfs}
\usepackage{array}
\usepackage{booktabs}
\usepackage{tabu}
\usepackage{dcolumn}
\usepackage{amsfonts}
\usepackage{float}
\usepackage{graphicx,color}
\usepackage[colorlinks={true}]{hyperref}
\hypersetup{colorlinks=true,linkcolor=red,citecolor=blue,urlcolor=blue}
\usepackage{bm}
\usepackage{subcaption}
\usepackage{pstricks}
\usepackage{braket}
\graphicspath{{images/}}
\usepackage{caption}
\expandafter\let\csname equation*\endcsname\relax
\expandafter\let\csname endequation*\endcsname\relax
\usepackage{mathtools}
\usepackage[utf8]{inputenc}
\usepackage[T1]{fontenc}
\usepackage{mathptmx}
\usepackage[markup=underlined]{changes}
\usepackage{cleveref}
\usepackage{etoolbox}
\usepackage{multirow}
\usepackage{orcidlink}
\usepackage{listings}
\usepackage{xcolor}

\begin{document} 

\title{Optimizing Continuous-Wave-Pumped Entanglement-based QKD in Noisy Environments} 

\author{Hashir Kuniyil\orcidlink{0000-0003-0338-1278}} 
\email{hkuniyil@hbku.edu.qa}
\affiliation{Qatar Centre for Quantum Computing, College of Science and Engineering, Hamad Bin Khalifa University, Doha, Qatar}

\author{Saif Al-Kuwari\orcidlink{0000-0002-4402-7710}}
\email{smalkuwari@hbku.edu.qa}
\affiliation{Qatar Centre for Quantum Computing, College of Science and Engineering, Hamad Bin Khalifa University,  Doha, Qatar}
\author{Asad Ali\orcidlink{0000-0001-9243-417X}}
\affiliation{Qatar Centre for Quantum Computing, College of Science and Engineering, Hamad Bin Khalifa University,  Doha, Qatar}
\author{Artur Czerwinski\orcidlink{0000-0003-0625-8339}}
\affiliation{Institute of Physics, Faculty of Physics, Astronomy and Informatics, Nicolaus Copernicus University in Torun, ul. Grudziadzka 5, 87-100 Torun, Poland}
\author{Syed M. Arslan\orcidlink{0000-0002-5820-6576}} 
\affiliation{Qatar Centre for Quantum Computing, College of Science and Engineering, Hamad Bin Khalifa University,  Doha, Qatar}

\date{\today}
\begin{abstract}
Quantum key distribution (QKD) has emerged as a promising solution to protect current cryptographic systems against the threat of quantum computers. As QKD transitions from laboratories to real-world applications, its implementation under various environmental conditions has become a pressing challenge. Major obstacles to practical QKD implementation are the loss of photons in the transmission media and the presence of extreme noise, which can severely limit long-range transmission. In this paper, we investigate the impact of extreme noise on QKD system parameters, including timing jitter, rate-dependent timing shifts, changes in effective detector dead time, and rate-dependent detection efficiency. Contrary to manufacturers' specifications, which assume these parameters to be constant, we demonstrate that these parameters exhibit significant variations in extreme noise conditions. We show that changes in these parameters play a key role in determining system performance in noisy environments. To address these nonidealities, we develop a model that adapts to detector-dependent timing distortions and recovery effects. In particular, our model is independent of source parameters and can be implemented using data from the detection unit. Our results show that the model enables reliable characterization and optimization of QKD performance under strong noise.
\end{abstract}

\keywords{Quantum key distribution, entanglement, coincidence measurement, avalanche photodetector, coincidence window optimization.}
\maketitle

\section{INTRODUCTION}
Quantum key distribution (QKD) is a method for securely transferring keys between multiple parties using quantum mechanics. QKD exploits quantum principles (including the no-cloning theorem) to detect eavesdropping in two-party cryptographic communication \cite{bennett1984proceedings,bennett2014quantum} and has been developed in various forms \cite{hwang2003quantum, lo2005decoy, ekert1992quantum, lo2012measurement, bennett1992quantum, yu2019sending}, including quantum-entanglement-based protocols \cite{ekert1992quantum, bennett1992quantum}, decoy-state prepare-and-measure schemes \cite{hwang2003quantum, lo2005decoy}, and more recently twin-field QKD protocols \cite{lucamarini2018overcoming, yu2019sending}. This system can further be extended to correct instrument-imposed loopholes with the assistance of device-independent QKD protocols \cite{nadlinger2022experimental, lo2012measurement}. 

These QKD methods have proven successful in both laboratory and practical settings, and significant efforts have been focused on extending the distance of individual quantum links using fiber-based communications \cite{li2019experimental, xu2020secure, neumann2022continuous} and free space connections, including terrestrial links \cite{peloso2009daylight, peng2005experimental, krvzivc2023towards} and satellite communications \cite{lim2021security, khatri2021spooky}. 

However, practical QKD faces major challenges, including higher noise levels and increased loss, which make the implementation of the protocol difficult. Unlike fiber-based connections, which are relatively immune to extreme noise, free-space links are more prone to intense noise interference \cite{krvzivc2023towards, mishra2022bbm92}. One way to reduce noise is spectral filtering, but its effectiveness is limited in extremely noisy environments \cite{krvzivc2023towards}. Temporal filtering offers additional noise filtering, taking advantage of the temporal correlation of the source's photons to remove unwanted photons \cite{peloso2009daylight,kupko2020tools,krvzivc2023towards}. Nevertheless, temporal filtering becomes ineffective in scenarios where noise photons are predominantly higher where the detector cannot function at this level of noise. Therefore, the aim is to maximize the performance within the functional level of noise of the detector.  

In Ref. \cite{neumann2021model}, authors achieved high performance by optimizing the coincidence window considering detector jitter, photon loss, and source photon's pair rate. Their experimental study considered a QKD system operating in the telecommunication band and used superconducting nanowire single-photon detectors (SNSPDs). However, they did not consider nonlinear detector imperfections that can arise at high detection rates. In addition, we observe a pronounced shift in the coincidence timing in the extreme-noise regime. This shift is rooted in photon arrival statistics and in rate-dependent recovery dynamics of silicon-based avalanche photodetectors. 

Similarly, the authors in \cite{stipvcevic2017advanced} conducted experiments to investigate variations in detector parameters, including dead time, timing jitter, detection timing shift, and afterpulsing effects as a function of detection count rate. They studied these parameters by sending pulsed laser light with well-controlled photon time separation to an actively quenched avalanche photodetector operated in a Geiger mode. Their results showed that these parameters are not constant and demonstrated a method to correct for these variations. This method is useful for mitigating deterioration factors in noisy detection. However, this approach is not applicable to continuous-wave (CW) pumped QKD systems, where photon arrival times are not externally controlled but governed by photon statistics.

In this paper, we performed an experiment to investigate and determine changes in detector parameters during high-photon detection events in CW-pumped systems. From the insight gained from this study, we develop a model to describe the coincidence measurement of the system and its dependence on variations in the timing jitter, detector dead time, and detection timing shift. 

To characterize and analyze the QKD system under high noise conditions, we used the spontaneous parametric down-conversion (SPDC) \cite{kwiat1995new,karan2020phase} process as a quantum source and an external broadband noise source features the noise photons. SPDC is a widely used quantum source for QKD realization, and the characteristics of broadband noise resemble noise expected in a practical system. Therefore, our method simulates a realistic scenario. Our findings from a single basis experimental study were extended to a two-basis theoretical model. This model enables the analysis of critical parameters of the QKD, including secure key rate (SKR), quantum bit error rate (QBER), and acceptable noise-photon levels within the coincidence window size of the system. Analyzing these parameters is crucial to improve system performance, particularly in extremely noisy environments.

The rest of this paper is organized as follows: In Section \ref{expMethods}, we discuss the effect of noise on coincidence measurements and present a simple experimental setup for this study. Part of this section also deals with the characterization of detector parameters, including timing jitter, afterpulsing, and detection timing shift. Section \ref{Theory} presents a theoretical analysis of a two-basis entanglement QKD system, which leads to the optimization of the coincidence measurement window. Finally, Section \ref{conclution} provides concluding remarks about our study.

\section{Experimental Analysis}\label{expMethods}
\subsection{Experimental Setup}
\begin{figure*}[htbp]
	\centering
	\includegraphics[width=\textwidth ]{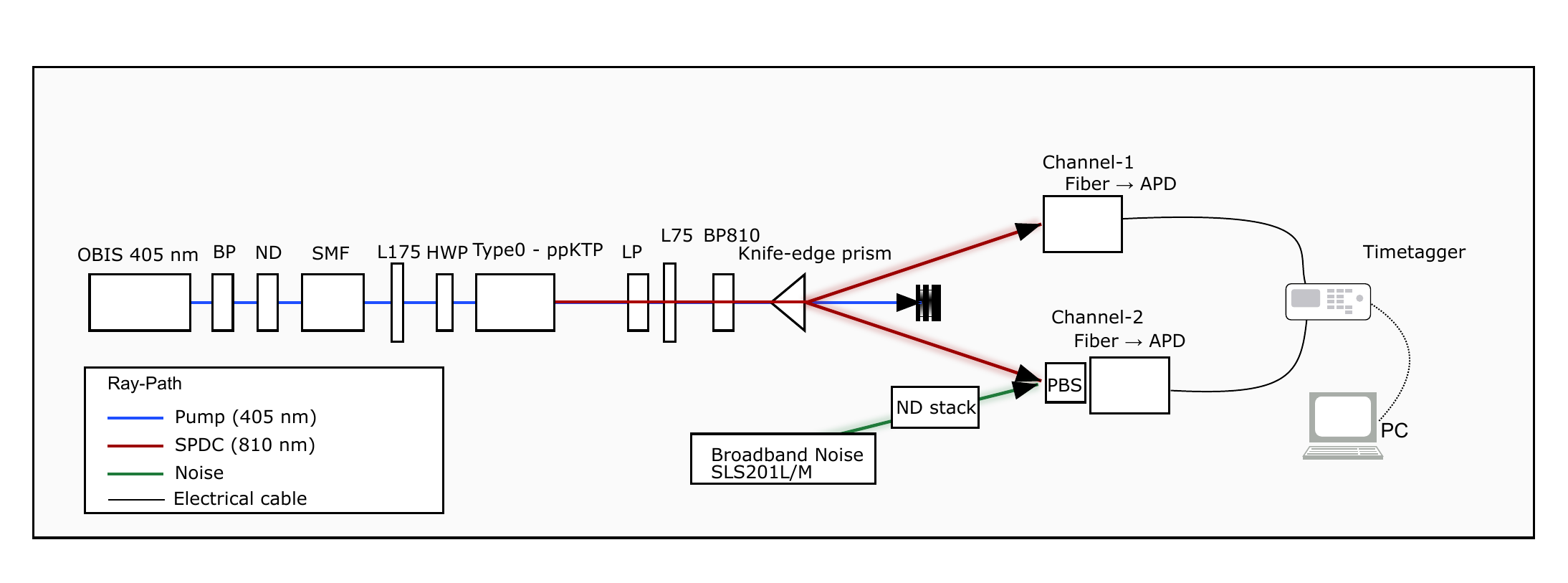}
	\caption{Experimental setup for coincidence profile characterization in noisy environments. BP - bandpass filter, LP - long pass filter, SMF - single mode optical fiber, L175 - lens with focal length 175 mm, L75- lens with 75mm.}
	\label{fig:1D experiment}
\end{figure*}
Our experimental setup is shown in Fig. \ref{fig:1D experiment}. We employ a single-frequency CW laser (OBIS, Coherent) operating at a central wavelength of 405 nm as the pump laser. The pump beam is spectrally cleaned using a bandpass filter with a bandwidth of 10 nm, followed by a variable neutral-density (ND) filter to control the pump power. The beam is then coupled into a single-mode optical fiber to ensure a well-defined spatial mode.

At the fiber output, the pump beam is focused into a type-0 periodically poled potassium titanyl phosphate (PPKTP) crystal using a plano-convex lens with a focal length of 175 mm (L175 in Fig. \ref{fig:1D experiment}). A half-wave plate is placed immediately after the fiber exit to optimize the pump polarization for efficient type-0 quasi-phase-matched SPDC. The crystal is aligned for collinear generation of photon pairs near the degenerate wavelength of 810 nm by tuning the temperature of the oven to $37^\circ$ Celsius.

After the PPKTP crystal, residual pump light is removed using a long-pass filter with a cutoff wavelength of 650 nm. The emerging collinear SPDC field, which exhibits a conical emission profile, is subsequently collimated using a lens with a focal length of 75 mm (L75 in Fig. \ref{fig:1D experiment}). A narrowband interference filter centered at 810 nm is then used to further reject any remaining pump photons and background light.

The collinear SPDC beam is spatially divided into two symmetric paths using a knife-edge prism positioned to bisect the transverse intensity profile of the SPDC beam, thereby separating the spatial mode into two channels. In both channels, photons are coupled into single-mode fibers and detected using identical fiber-coupled avalanche photo detectors (APDs, Excelitas SPCM-AQRH-14-FC). The Salient features of this detector are summarized in Table~\ref{table:1}).

The two detectors collect The photon arrival times are recorded using a Swabian Instruments Time Tagger Ultra (timing resolution 8 ps), but histogram bin width we used is 85 ps. With this arrangement, The noise-free coincidence rate of the channel-1 and channel-2 is calculated to 6321 counts per second (cps) when the measurement window selected to be 16.915 ns, and its heralding efficiency is 3.2 $\%$ throughout the coincidence measurement. In Channel 2, controlled noise is deliberately introduced by mixing the SPDC channel-2 photons with light from a broadband source (Thorlabs SLS201L/M). This noise source is a Tungsten-Halogen Light Source. Since this source exhibits thermal light characteristics with broad spectral width, and there can simulate sunlight. A polarizing beam splitter is used to efficiently mix the noise and channel-2 signal. Stack of neutral-density filters are placed in the noise arm that allows well control the noise intensity. This configuration enables systematic investigation of detector response under varying noise conditions while maintaining identical detection hardware in both channels.

In our experiment, only APD at channel-2 received noise (channel-1 is shielded from noise). Therefore, the measured output from channel-2 is a mixture of signal and noise. These two detectors are connected to a time tagger to record the time of arrival of individual photons received at both detectors. By analyzing the joint time of arrivals of the photons at the detectors, we could obtain a temporal coincidence histogram as shown in Fig.\ref{fig:start-stop_excelitas} (the python code to achieve the coincidence measurement is given in the appendix. \ref{app_A}). This temporal coincidence measurement is a basis for most of the discrete variable QKD experiments. In our setup experiment, idler photons travel $\sim 0.74$ m extra optical path (equivalently, a $\sim2.46$ ns in time domain); therefore, the coincidence histogram has an offset from zero point as evident from Fig. \ref{fig:start-stop_excelitas}. 
\begin{figure*}
\centering
\begin{subfigure}{0.5\textwidth}
  \centering
  \includegraphics[width=1\linewidth]{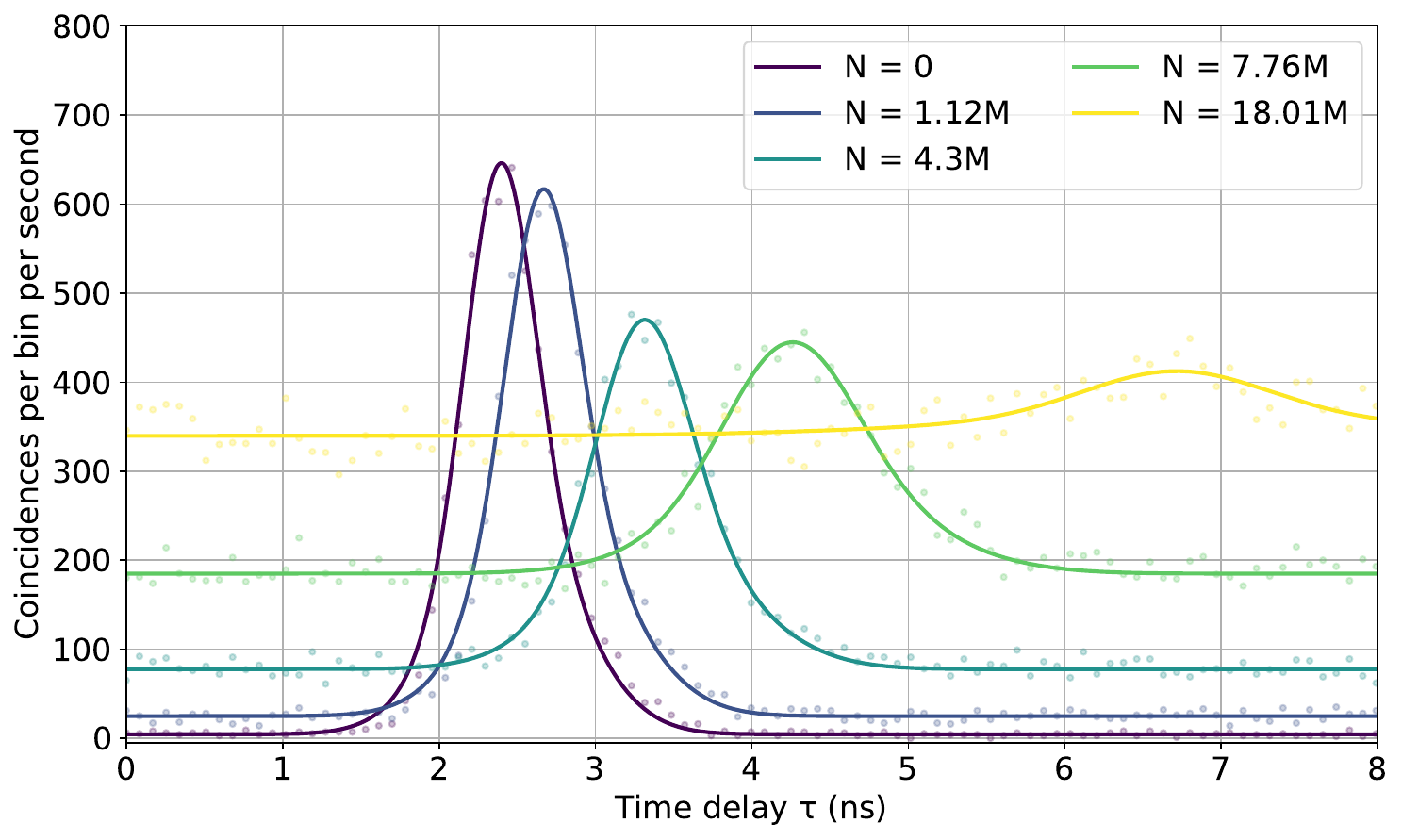}
  \caption{Coincidence histograms versus delay for different detected photon rates using the SPCM-AQRH-14-FC} detector
  \label{fig:start-stop_excelitas}
\end{subfigure}%
\begin{subfigure}{.5\textwidth}
  \centering
  \includegraphics[width=1\linewidth]{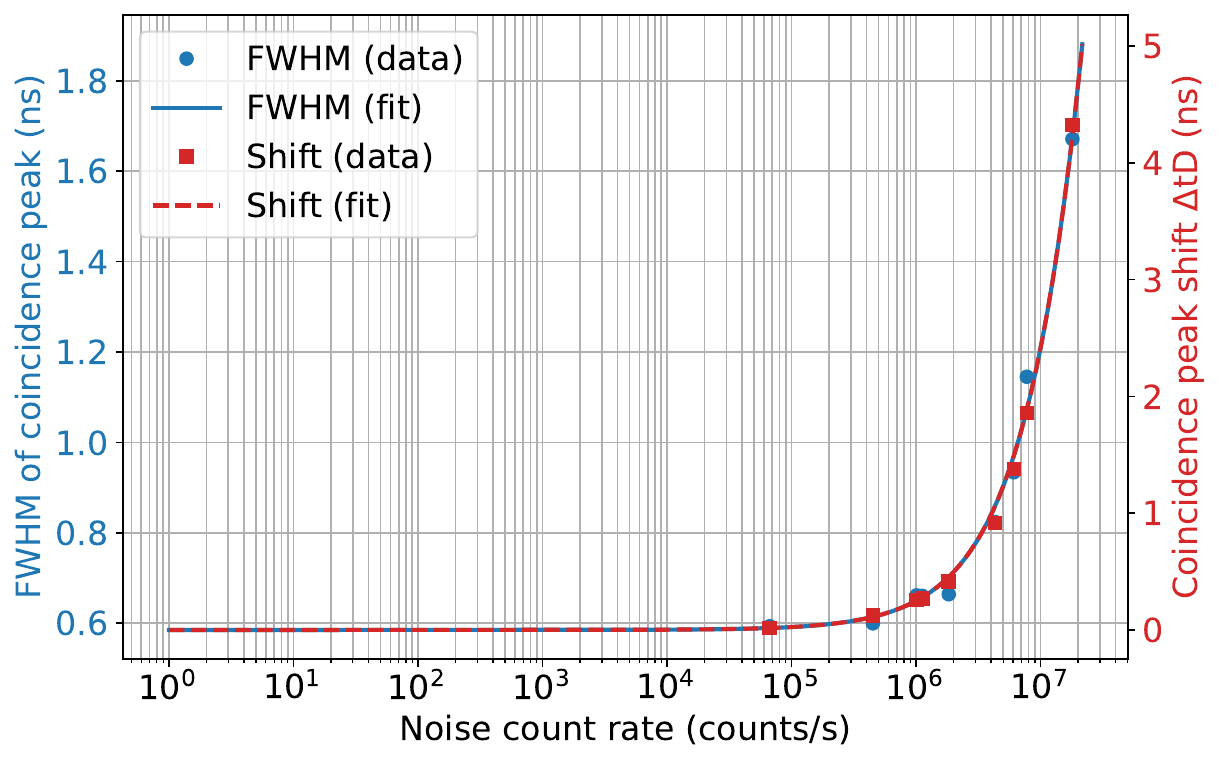}
  \caption{Coincidence peak full width at half maximum (FWHM, left axis) and peak shift relative to the low-noise condition (right axis) as a function of injected noise count rate. Solid and dashed lines show fits to a saturating recovery model with a shared characteristic rate 
$R_0 = 1.43\times10^8$, indicating a common rate-dependent detector recovery mechanism underlying both effects.}
  \label{fig:sub2}
\end{subfigure}
\caption{Effects of noise on the performance of single-photon detectors.} 
\label{fig:test}
\end{figure*}

\subsection{Coincidence Measurement in Noisy Settings}\label{coinMeasure}
In a typical QKD protocol (e.g., BBM92 protocol \cite{bennett1992quantum}), Alice and Bob each require a minimum of two detectors to distinguish orthogonal quantum states. This requirement stems from the protocol's security mechanism, which relies on the quantum nature of a maximally entangled state and necessitates measurements in two bases. Implementing the protocol requires examining the coincidence counts of measurement events between Alice and Bob using two detector combinations. Therefore, single-basis experimental settings are not sufficient for a complete protocol implementation.
However, to study the factors affecting SKR and QBER and their performance changes due to operation in a noisy environment, analyzing a single coincidence channel is sufficient to isolate the dominant timing and noise effects. One key difference is that the photon coincidence rate obtained by a single basis system will be twice that of two basis measurements. Additionally, it is important to note that by considering only one detector at Alice and Bob locations, we are ignoring the polarization error from the basis setting. This is justifiable as this factor will contribute less than $1\%$ of the total errors \cite{anwar2021entangled}. Additionally, this only leads to a multiplicative change in the overall error rate. 

Fig.~\ref{fig:start-stop_excelitas} shows a plot generated under various noise conditions (the actual noise values are indicated in the figure). In our analysis, the number of coincidences is determined by accumulating the number of coincidence counts in every second. By compiling the experimental coincidence values, we can obtain a full coincidence counts' distribution that provides important information about the system operating in a noise setup. For APD-based detection, the coincidence-peak width is primarily determined by detector timing jitter (convolved across both detectors). Our detectors have timing jitter of 350 ps (according to the manufacturer), which we verify by measuring the full-width at half-maximum (FWHM) of the coincidence distribution in a low noise condition. We observed four distinct changes in the coincidence counts as a function of increasing detection counts (driven by noise in this case). 
\begin{enumerate}
\item A systematic shift of the coincidence peak, which becomes evident above $\sim$1 Mcps and increases with count rate.
\item Broadening of the coincidence peak (FWHM increase), also evident above $\sim$1 Mcps.
\item An increase in the coincidence baseline due to accidental coincidences.
\item A reduction in peak amplitude at high count rates, consistent with rate-dependent efficiency degradation.
\end{enumerate}
The coincidence peak shift originates from rate-dependent distortion of the single-detector timing response. As noise is injected into one detection channel, the inter-arrival time distribution becomes increasingly truncated by detector dead time, leading to a delayed and asymmetric temporal response. Since the coincidence histogram represents a convolution of the two detector response functions, the asymmetric distortion of one channel shifts the location of maximum temporal overlap, resulting in a systematic displacement of the coincidence peak.
If we account loss of photon during the transmission, we can estimate the photon detection rate on the detector using
\begin{align}
    S = B\eta,
    \label{eq:singles value}
\end{align}
where $B$ is the number of pairs generated in the crystal, $\eta$ is the detection efficiency. We assume transmission loss, and other loss parameters negligible. With this, coincidence detection probability of photons at channel-1 and channel-2 can be estimated following
\begin{align}
    C = B\eta_1\eta_2.
\end{align}
Where $\eta_1$ and $\eta_2$ are the quantum efficiencies of detectors at channel-1 and Channel-2, respectively. The detection events registered by the detectors also have noise counts, such as the dark count of the APDs ($DC$) and stray photons entered in the detectors ($N$). which can be modeled by modifying Eq.~(\ref{eq:singles value}) into

\begin{align}
    S_1 = B\eta_1+N_1+DC_1\\
    S_2 = B\eta_2+N_2+DC_2,
\end{align}
where $N_1$ and $N_2$ are the count of noise entered in channel-1 and channel-2, $DC_1$ and $DC_2$ are the dark count rates of detector-$1$ and detector-$2$, respectively. In the measurement, there is a random probability that two uncorrelated photon events fall within the same measurement time window. This is known as accidental coincidence counts. These events are the primary source of noise in coincidence-based measurements. The rate of accidental coincidence increases with the noise photons following the relation
\begin{equation}
    N_{AC} = S_1S_2\tau_{cw},
    \label{eq:coincidence_events}
\end{equation}
where $\tau_{cw}$ is the selected coincidence measurement time window. In experiments, the accidental coincidence can be estimated by measuring y-axis offset of start-stop histogram (see Fig.~\ref{fig:start-stop_excelitas}). Unlike genuine coincidences, accidental coincidences produce a nearly time-independent background in the start–stop histogram. As the dark count of a detector is constant and is known from the manufacturer, the environment-impacted change in the detection events is the only varying parameter.

Typically, the coincidence measurements are theoretically approximated using a normal distribution of the form \cite{sedziak2020tomography,czerwinski2021phase}
\begin{align}
    j(\tau) = \frac{1}{\sigma\sqrt{2\pi}}\exp\left(-\frac{\tau^2}{2 \sigma^2}\right),
    \label{eq:detector_fn}
\end{align}
where $\sigma$ is the detector timing jitter and $\tau$ is the measurement time. This method can be applied in numerical simulations to investigate quantum state tomography of entangled qubits with measurements affected by time uncertainty \cite{czerwinski2022quantum}. However, this approximation fails when systems operate in high noise regimes. Also, this approximation does not allow one to study individual detector characteristics from the coincidence histogram. In our experiment, where only one detector receives noise-induced clicks, the probabilistic distribution of the coincidence counts does not show a clean Gaussian profile in the high noise regime, rendering a normal distribution that does not fit our collected data. Neumann \emph{et al.} \cite{neumann2021model} investigated coincidence counts using the normal distribution function for various coincidence values, accounting for variations in the Full Width at Half Maximum (FWHM) and shifts caused by dispersion in optical fibers. Using the detector function of the form of Eq. (\ref{eq:detector_fn}), they characterized the changes in the detector parameters using linear functions. In contrast, our observations reveal a polynomial growth in FWHM (see Fig~\ref{fig:test}b) and a shift in the coincidence time. This discrepancy arises because previous work assumed equal photon pair detection rates across all detectors in a low-noise scenario. Consequently, in their study, FWHM changes and coincidence peak shifts appeared linear. 

Our results show external photons can induce the shift in the detection timing. In particular, this becomes evident when detection rates are imbalanced. To efficiently address this, it is therefore necessary to introduce a model that could cater changes in the individual detectors in the coincidence profile. By incorporating noise-induced variation parameter, one could detect changes in the individual detectors using the formulation of
\begin{align}
    G= N_{AC}+B\eta_1\eta_2(1-t_{dA}S_1\eta_1) A_i +\nonumber\\ B\eta_1\eta_2(1-t_{dB}S_2\eta_2) A_j
\end{align}
\begin{align}
    A_{i} &=e^{W_i\tau}  & \text{for $\tau>C$}\nonumber\\
    A_{j} &=e^{-W_j\tau}  & \text{for $\tau<C$},
    \label{eq:correction}
\end{align}
where $W_{i(j)}$ is a coefficient used to accommodate the change in the width of coincidence counts, $t_d$ is the dead time of the detector and $C$ is the centroid value of the coincidence histogram. $1/W$ will provide the FWHM in the time domain. By incorporating detector function in Eq. (\ref{eq:detector_fn}) with $A_{i}$ and $A_{j}$, we can obtain a complete coincidence histogram function of the form:
\begin{align}
    C(\tau, W_1, W_2, t_D)  = S_1S_2\tau_{cw}+B\eta_1\eta_2(1-t_{dA}S_1\eta_1)\nonumber \\ 
    e^{W_1\left(W_1\sigma_1^2/2+\tau -t_D\right)} \mathrm{erfc}\left[\frac{W_1 \sigma_1^2+\tau- t_D}{\sqrt{2}\sigma_1}\right]\nonumber\\
    +B\eta_1\eta_2(1-t_{dB}S_2 )e^{W_2\left(W_2\sigma_2^2/2-\tau+t_D\right)} \mathrm{erfc}\left[\frac{W_2 \sigma_i^2-\tau+t_D}{\sqrt{2}\sigma_2}\right], 
    \label{eq:E8}
\end{align}
where $\sigma_1$ and $\sigma_2$ are the timing uncertainty (jitter) of the detectors used at channel-1 and channel-2, respectively, $t_D$ is the shift of centroid point of coincidence histogram from the origin, and ``$\mathrm{erfc}$'' represents the complementary error function. Detector timing jitter is predominantly responsible for the width of the coincidence counts. If the temporal coherence of the collected photons is significantly greater than the jitter of the detection system, the measured spectral profile will reflect the true spectral characteristics of the photons more closely \cite{clausen2014source, srivathsan2013narrow}.  The coherence time of our photon pairs is about 60 fs (corresponding to the spectral bandwidth of approximately 16 nm, this value is taken from Ref. \cite{steinlechner2014efficient}), which is negligible and overshadowed by the timing jitter of the detectors. Therefore, timing uncertainty due to spectral width of the SPDC photons could be omitted. Additionally, the temporal resolution of the time tagger also contributes to the width of coincidence counts if it is greater than the timing jitter. Since the temporal resolution of our time tagger is 8 ps, since this is much less than $\sigma$, we can omit this factor as well. In a noise-free operation of the detectors, we can approximate $W_{1(2)} = 1/\sigma_{1(2)}$. As noise increases, we adjusted the value of $W_{s(i)}$ to fit with the experimental data.

In Fig.~ \ref{fig:start-stop_excelitas}, we used Eq. (\ref{eq:E8}) to fit the experimental result. The change in the FWHM in different noise settings is extracted and plotted in Fig.~\ref{fig:sub2}. The shift of the start-stop histogram also extracted from Fig.~ \ref{fig:start-stop_excelitas} and plotted in Fig.~\ref{fig:sub2}.
\subsection{Detection timing shift and FWHM widening in start-stop histogram}
Since we saturate only one detector, the coincidence peak measures the mean relative delay between channel-1 and channel-2. At low count rates, each avalanche is fully quenched and the detector fully recovers between detection events. The discriminator threshold is crossed at a fixed point in the avalanche buildup. In this case, the detection latency is deterministic  with fixed jitter. Mathematically, detection time is:
\begin{align}
    t_{\det} = t_{\text{true}} + \delta_0 + \epsilon
\end{align}
where $\delta_0$ is the fixed latency, and $\epsilon $ is the stationary jitter (Gaussian). As a result in coincidence, we observe stable peak position, constant width (constant FWHM) of start-stop coincidence measure, maximum peak height, and low accidental background. When we increase the noise rate in detector-2, this detector rarely fully recovers. Many photons arrive during the recovery tail. Consequently, effective bias voltage is lower than nominal. Avalanche growth is slower. Therefore, detection latency becomes state-dependent. The detection delay now depends on the time since the last avalanche: $\delta=\delta(t_{since\space last})$. This function increases as recovery worsens, and saturates at high rate. In our case detector-1 is stable and detector-2 develops a positive mean latency shift in a count regime:
\begin{align}
    \langle \delta_2(R) \rangle > \delta_0
\end{align}
Thus the rate-dependent delay we observe is:
\begin{align}
    t_D(R) = t_{D0} + \langle \delta_2(R)\rangle
\end{align}
Because recovery is related to Poisson statistics, the probability that at least one photon appears in the time "t" is given by
\begin{align}
    P_{\ge1}(R) = 1-e^{-Rt},
    \label{eq:13}
\end{align} 
where R is the photon rate. Due to characteristic recovery time of the detector, it cannot detect photons if $t<\tau_0$, where $\tau_0$ is the characteristic recovery time of the detector. Thus, Eq. \ref{eq:13} modify into
\begin{align}
    P_{\ge1}(R) = 1-e^{-R\tau_0}
    \label{eq:13}
\end{align}
The coincidence peak shift we observed should follow this behavior with intrinsic delay $t_{D0}$ incorporated in it. Thus, the shift we observed can be estimated using
\begin{align}
    t_D(R) = t_{D0} + A\left(1 - e^{-R/R_0}\right),
    \label{Eq15}
\end{align}
where "A" is the amplitude of this effect, $R_0 = 1/\tau_0$ is the characteristic rate. We also observed widening of start-stop histogram. This is because not all detections are delayed by the same amount. The recovery state varies randomly event-to-event, so the detection latency distribution broadens. Thus, detector-2 has larger variance in detection time. In low count rate measurement the Coincidence width obeys
\begin{align}
    \sigma_{\text{coin}}^{2} \approx \sigma_{1}^{2} + \sigma_{2}^{2}.
\end{align}
Since only $\sigma_2$ increases, FWHM increases, and peak becomes asymmetric. This extra jitter saturates at high rate. Since both peak shift and increase in FWHM follow same recovery dynamics, we expect both follows the trend. Following this, we can observe widening of the FWHM of the start-stop coincidence profile by following
\begin{align}
    FWHM(R) = W_0 + \Delta W \left(1 - e^{-R/R_0}\right).
\end{align}
The experimentally extracted FWHM and peak shift are shown in Fig.~\ref{fig:sub2}. Both are well fitted by a saturating recovery model with a shared characteristic rate, supporting the hypothesis of a common recovery-driven mechanism.
This summary, injecting noise photons into one detector introduces a rate-dependent modification of the detector recovery dynamics. At high count rates, photons are detected before full recovery, leading to both an increased mean detection latency and enhanced variability in the detection time. In coincidence measurements, this manifests as a shift of the coincidence peak, a reduction in peak amplitude due to reduced effective detection efficiency, an increase in the accidental background, and a broadening of the coincidence peak. Both the peak shift and the FWHM exhibit a saturating dependence on the noise count rate with a shared characteristic rate, indicating a common physical origin in detector recovery effects.
\begin{figure*}[ht]
	\centering
	\includegraphics[width=\textwidth]{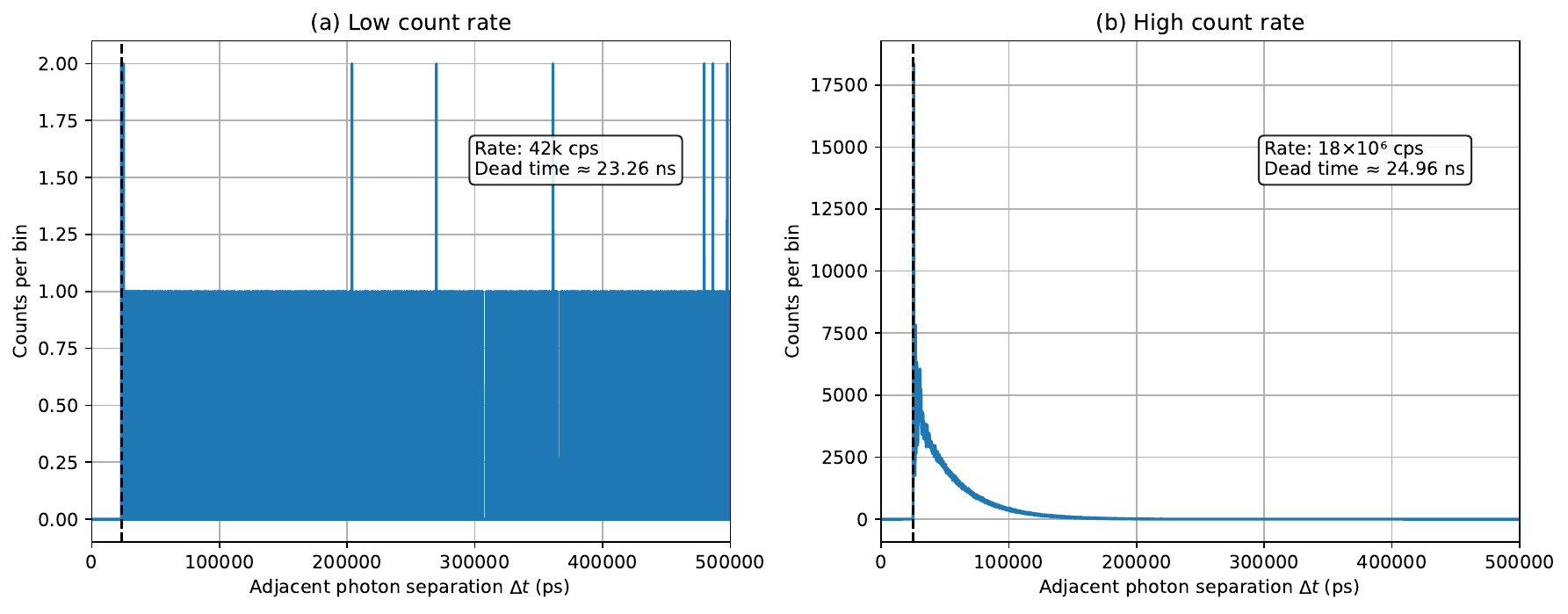}
	\caption{Single-detector inter-arrival time distributions at low and high count rates. Inter-arrival time distributions of adjacent photon detection events recorded from a single avalanche photodiode operated at different photon count rates. Panel (a) shows the distribution measured at a low count rate ($\approx$ 42 kcounts/s), while panel (b) shows the distribution measured at a high count rate ($\approx$ 18 Mcounts/s). The horizontal axis represents the time separation $\Delta t$ between successive detection events, and the vertical axis shows the number of events per time bin. Dashed vertical lines indicate the minimum detectable photon separation, corresponding to the effective detector dead time extracted from each dataset. At low count rates, the distribution is approximately uniform beyond the dead time, indicating that the detector fully recovers between detection events. At high count rates, a pronounced accumulation of events near the dead time threshold is observed, demonstrating rate-dependent detector recovery dynamics that lead to increased effective dead time and enhanced detection immediately after recovery.}
	\label{fig3}
\end{figure*}
\begin{table}[h!]
\centering
\begin{tabular}{||c c||} 
 \hline
Parameter &  SPCM-AQRH-14-FC \\ 
\hline
Dark count & 39 cps\\
maximum count & 38.1 MHz \\ 
dead time & 23 ns (@ < 1MHz)\\ 
timing jitter & 350 ps \\ 
efficiency@ 780 nm & 59 \% \\ 
active area & 180 um \\ 

 \hline
\end{tabular}
\caption{Company provided performance data of SPCM-AQRH-14-FC single photodetector module}
\label{table:1}
\end{table}

\subsection{Characterization of Detector Parameters}\label{detectorChara}
The SPCM-AQRH-14 detector is a commonly used single-photon detection module in QKD \cite{stipvcevic2017advanced, sajeed2015security, ursin2007entanglement}. This is an actively quenched Geiger mode detector and a nonparalyzable detector; after a detection event, it remains insensitive for a fixed dead time $t_d$. The manufacturer-provided parameters of this detector are summarized in Table.~\ref{table:1}. To demonstrate that the effective recovery time of the avalanche photodiode (APD) increases with photon count rate, we first characterized the timing response of a single detector operated under varying detection rates. From the time-tagged photon arrival records, we evaluated the temporal separation between successive detection events, defined as
\begin{align}
    \Delta t = t_{n+1} - t_n.
\end{align}
Histograms of $\Delta t$ were constructed using a timing resolution of $8$ ps, as shown in Fig. \ref{fig3}. At low detection rates, the distribution of $\Delta t$ is approximately flat over the chosen time window, indicating a uniform probability of detecting photons at different separations (though minimum detection time is found out to be 22 ns, exactly the dead time value provided by the manufacturer). This behavior reflects the fact that the detector has sufficient time to fully recover between successive detection events. As the photon count rate increases, the probability of observing short inter-arrival times increases noticeably, with a pronounced accumulation of events near the minimum detectable separation, as shown in Fig. \ref{fig3}(b).

We interpret this minimum photon separation as the effective detector dead time, defined as the shortest interval following a detection during which the detector can register a subsequent photon. With further increases in the photon count rate, detection events at this minimum separation occur with increasing frequency, providing direct evidence that the effective dead time—or, more generally, the detector recovery dynamics—becomes rate dependent as the detector is driven toward saturation.

\begin{figure}[ht]
	\centering
	\includegraphics[width=0.5\textwidth]{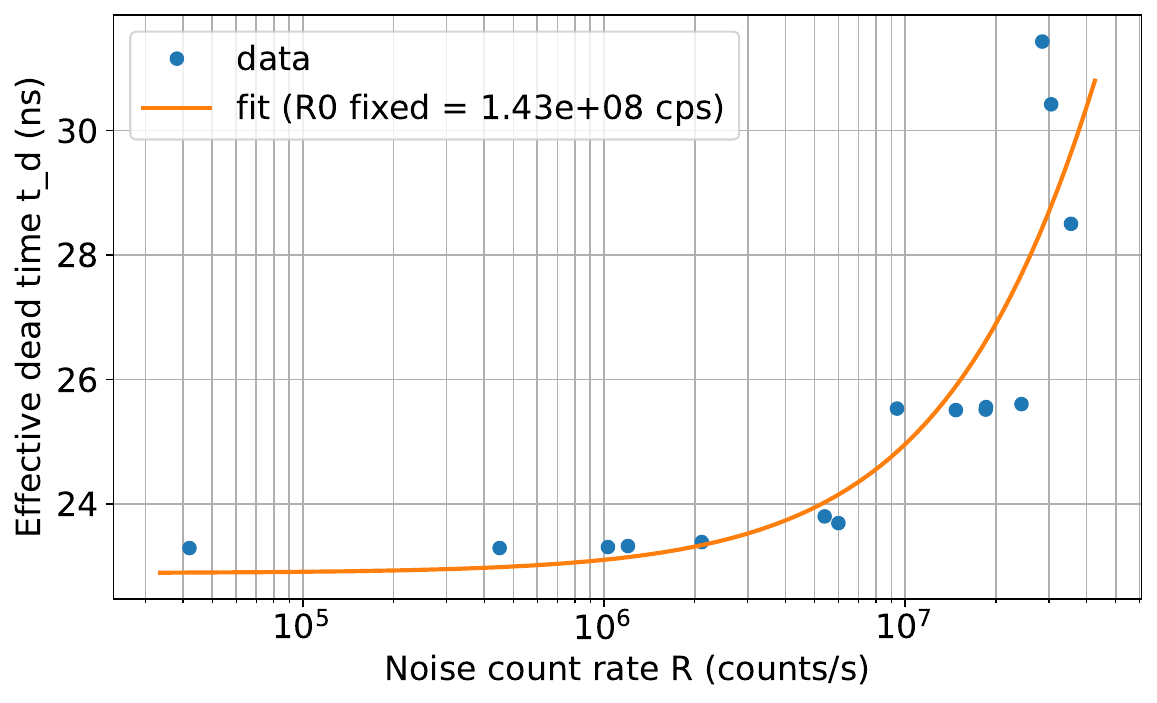}
	\caption{Effective dead time as a function of photon count rate.}
	\label{fig4A}
\end{figure}

The extracted effective dead time as a function of the photon count rate is summarized in Fig. \ref{fig4A}. To describe this behavior, the data were fitted using a saturating recovery model of the form
\begin{align}
    t_d^{eff}(R) = t_{d0} + \Delta\bigl(1 - e^{-R/R_0}\bigr)
\end{align}
where $t_{d0}$ is the initial (or baseline) delay time, $\Delta$ is the maximum additional delay induced by recovery effects at high count rates, and $R_0$ is a characteristic recovery rate. The fit captures the observed increase and saturation of the effective dead time with increasing photon flux.

To establish that these recovery dynamics are responsible for the shift and broadening observed in the coincidence histograms, we compare the extracted single-detector recovery behavior with two-detector coincidence measurements. As shown in Fig. \ref{fig:sub2}, the coincidence peak position shifts systematically with increasing noise count rate, while Fig. Fig. \ref{fig:sub2} also shows a concomitant increase in the coincidence peak full width at half maximum (FWHM). Both quantities are well described by the same functional dependence on the count rate, with the coincidence peak shift following Eq. \ref{Eq15}. and the FWHM following an analogous saturating form.
Importantly, the characteristic rate $R_0$ extracted independently from the single-detector dead time analysis (Fig. \ref{fig3}), the coincidence peak shift, and the coincidence peak broadening (Fig. \ref{fig:sub2}) is consistent across all measurements. This quantitative agreement demonstrates that the observed coincidence peak shift and broadening originate from the same underlying physical mechanism: rate-dependent detector recovery dynamics in the APD. The consistency between single-detector timing statistics and two-detector coincidence measurements confirms that detector recovery effects, rather than optical or electronic artifacts, govern the observed distortions in the coincidence profiles.
\section{Theoretical Analysis}\label{Theory}
Given the trend of coincidence counts in the presence of extreme noise, one should investigate its influence on the SKR and QBER. Typically, a coincidence window (a chosen measurement time window to observe coincidence counts) is optimized to achieve a maximum SKR with a manageable QBER. Due to the Gaussian shape of coincidence counts, optimal coincidence window selection is crucial for improved QKD performance. Additionally, the change in the shape of the coincidence histogram due to extreme noise should also be considered when choosing the coincidence window. 

In this section, we theoretically investigate the factors to consider when determining the coincidence window in a QKD system. We first investigate influencing factors in a single-basis measurement setting and then study effects on a complete two-basis QKD system.

\subsection{Coincidence Measurement in a Single Basis System}
Eq. (\ref{eq:E8}) provides a complete measurement result, including the noise effect on the coincidence profile. As a result, the total coincidence values within the chosen coincidence window can be estimated by carrying out the integration
\begin{align}
    E = \int_{a}^{b} C(\tau, W_1, W_2, t_D) d\tau.
    \label{eq:E9}
\end{align}
Where $E$ is the total number of coincidence counts with start-bin $a$ and stop-bin $b$. When we determine the start-bin and stop-bin values and compute the total coincidence counts, $E$, using Eq.~(\ref{eq:E9}), the result will contain all coincidence counts including coincidence counts from generated photon pairs and accidental coincidence counts. Notice that our experiment is carried out with time tagger resolution of 85 ps. Because the experimental histogram is reported in counts per bin per second, we evaluate the total coincidence rate in a window by summing over bins (equivalently, integrating the corresponding rate density over$\tau_{bin}$) in Eq. (\ref{eq:E9}). When the detectors operate in a low noise regime, the share of accidental coincidence in the computed ``C'' value is low. In this case, it is desirable to select a wide coincidence window.  However, in extreme noise events, the share of accidental coincidence is higher, and the target in this case would be to minimize the accidental coincidences so that QBER can be lowered. The QBER and accidental coincidence are directly proportional and hence, in a single basis system, minimizing the ratio of accidental coincidence with total coincidence would be one target parameter(accidental coincidence fraction, $f_{acc}$) that could approximate QBER measurement, i.e.
\begin{align}
    f_{acc} = \frac{N_{acc}}{E}.
    \label{eq:E10}
\end{align}
where $N_{acc}$ is the estimate of accidental coincidence, experimentally we can estimate it by $N_{acc} = A_{cci}\times N_{bin} T$, where $A_{cci}$ accidental coincidence per bin, $N_{bin}$ ($N_{bin} = t_{CW}/\Delta\tau_{bin}$) is the number of bins selected, and $T$ is the experimental duration. From our measured and fitted curve we can estimate $f_{acc}$:
\begin{align}
    f_{acc} = \frac{A_{cci}\times N_{bin}}{\Sigma_{bin}y_{bin}}.
    \label{eq:E10}
\end{align}
In QKD, this is closely related to an error contribution, but whether it equals the QBER depends on the encoding/basis. If accidentals are uncorrelated with the correct outcome, they typically contribute $~50\% $ error among the accidental subset. which means $QBER_{acc} = f_{acc}/2$.

Since the width of the coincidence histogram increases with noise plus the its timing shift essentially require us to optimize the coincidence window. It is clear from our experimental results that fixed choice of coincidence window will cause loss of useful measurement and its wrong choice would lead to noise induced high QBER.

\subsection{QKD Performance Under Noisy Conditions}
\begin{figure}
	\centering
	\includegraphics[width=0.5\textwidth ]{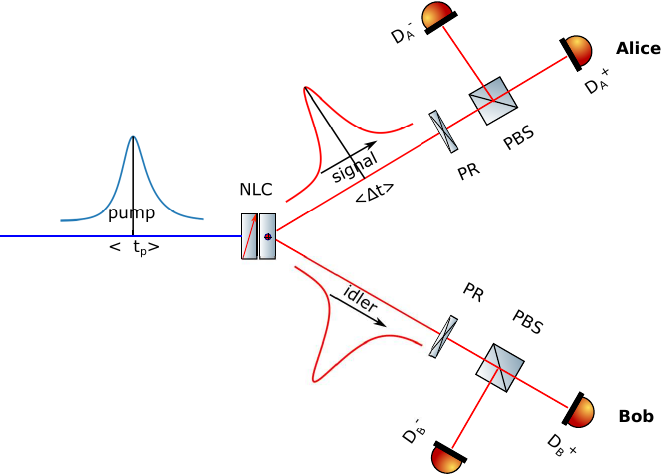}
	\caption{Typical experimental setup used to implement BBM92 QKD protocol. NLC - nonlinear crystal, PR - polarization rotator, PBS - polarizing beam splitter.}
	\label{fig:fig4}
\end{figure}
To analyze the changes in the complete QKD system due to noise, measurements in at least two bases are necessary. This requires Alice and Bob to each have two detectors, with the capability to randomly select the basis from a two conjugate basis set. To satisfy this condition, the quantum source in entanglement-based QKD must generate Bell states, which can be defined as:
\begin{equation}
    \ket{\psi} = \alpha \ket{H_A H_B} +\beta \ket{V_A V_B}, 
    \label{eq:E1}
\end{equation} 
where quantum states $\ket{H}$ and $\ket{V}$ represent horizontal and vertical polarization, respectively, while the subscripts $A$ and $B$ denote Alice and Bob photons, respectively. The parameters $\alpha$ and $\beta$ represent amplitudes with probabilities given by $|\alpha|^2$ and $\beta|^2$ , respectively, such that $|\alpha|^2+|\beta|^2 = 1$. In a balanced system, $\alpha = \beta = 1/\sqrt{2}$. Thus, the probabilities for $\ket{HH}$ and $\ket{VV}$ are equal. In the implementation of a typical QKD protocol, bases are randomly set into two states (in BBM92 bases $45^\circ$ and $0^\circ$ are randomly set). Statistically, half of the total photon pairs generated will undergo measurement in the $45^0$ basis, while the remaining half will be measured in the $0^0$ basis. Therefore, there must be at least two coincidence measurements, one between detector $D_A^-$ and $D_B^-$ and the other between $D_A^+$ and $D_B^+$ (see Fig. \ref{fig:fig4}). To understand the total coincidences in this case, we can modify Eq. (\ref{eq:E9}) and obtain 
\begin{align}
    C  = \int_{A^+}^{B^+}C^+(\tau, t^+_\Delta, t^+_D)+\int_{A^-}^{B^-}C^{-}(\tau, t^-_\Delta, t^-_D)d\tau,
    \label{eq:twobasis_start_stop}
\end{align}
where `$+$' and `$-$' indicate the positions of the detectors in the experiment as depicted in Fig. \ref{fig:fig4}. Eq. (\ref{eq:twobasis_start_stop}) will provide complete coincidence values generated with the flexibility to select the coincidence window. Notice that the brightness ,$B$, in Eq.~(\ref{eq:E8}) will turn to $B/2$ as total generated photon pairs must be equally shared between detectors. With this, the QBER ($Q$) of the system can be redefined as
\begin{align}
   Q &= 1/2(A^++A^-)/C\\
   &= 1/2(f^++f^-),
   \label{eq:QBER}
\end{align}
where $A^+$ is the accidental coincidence between the '$+$' detectors, and $A^-$ is the accidental coincidence between '$-$' detectors. We could evaluate $A^+$ and $A^-$ by using the following
\begin{align}
    A^{+(-)} &= S_A^{+(-)}(1-t_{dA^{+(-)}}S_A^{+(-)})S_B^{+(-)}(1-t_{dB^{+(-)}}S_B^{+(-)})\tau_{cw}\\
    &=A_{acci}^{+(-)}N_{bin}T.
\end{align}
When implementing the protocol, Alice and Bob randomly switch their measurement basis between $0^{\circ}$ and $45^{\circ}$. Next, they publicly reveal their basis choices. Due to random selection, they have only 50\% probability of choosing the correct basis. Only compatible basis choices are considered for protocol implementation, hence, 50\% of the coincidence measurements are discarded. Therefore, the total achievable secure key rate ($R$) in this procedure could be given by \cite{ma2007quantum}
\begin{align}
    R = \frac{1}{2}C[1-f(Q_{bit})H_2(Q_{bit})-H_2(Q_{phi})],
    \label{eq:16}
\end{align}
where $H_2 (x)$ is the binary entropy function, defined by
\begin{align}
    H_2 (x) = -x \log_2x-(1-x) \log_2(1-x),
\end{align}
where $Q_{bit}$ is the error in bit streams, $Q_{phi}$ is the phase error between $\ket{H}$ and $\ket{V}$ polarization states, and $f(Q_{bit})$ accounts for the additional bits sacrificed during the reconciliation process to ensure Alice and Bob share an identical key sequence. Reconciliation is a crucial step in QKD, where classical error correction protocols are applied to correct discrepancies in the sifted key while minimizing information leakage to a potential eavesdropper. As a realistic value of $f(Q_{bit})\leq 1.1$ \cite{elkouss2010information}, we considered the maximum limit while executing the model. Due to the symmetry of the basis choice, we can consider $Q_{phi} = Q_{bit} = Q$. Incorporating these, finally we turn Eq.~(\ref{eq:16}) into
\begin{align}
    R = \frac{1}{2}C[1-2.1H_2(Q)].
    \label{eq:skr}
\end{align}
Using Eq. (\ref{eq:skr}) and (\ref{eq:QBER}), we examine SKR and QBER in noisy scenarios. With a pair rate ($B$) of 500K cps distributed equally at the Alice and Bob sites, Fig. \ref{fig:fig5} illustrates the trend of SKR and QBER in various noise levels for different start-bin and stop-bin choices. In this demonstration, we considered equal noise distributed to all detectors. The result shows that a QBER threshold of $Q = 0.102$  (indicated by the dark dotted line in Fig. \ref{fig:fig5}) is established, above which no secure key can be generated. At this threshold, the secure key rate becomes zero ($R = 0$). Therefore, achieving $Q<0.102$ is a target for the implementation of QKD. To reach this threshold, different start and stop bin choices tolerate different levels of noise. For example, when the start bin = 3 ns and the stop bin = 6 ns, the system tolerates 26 Mcps noise counts before reaching $R = 0$. However, it accepts 17 Mcps noise counts when the start of the bin = 1 ns and the stop bin = 8 ns, despite the larger total coincidence window. When the start bin is set near the centroid of the coincidence window (the start bin = 4.5 ns) and the stop bin = 7.5 ns, the system shows reduced resistance to noise, tolerating only 13.5 Mcps. 

\subsection{Optimization of the Coincidence Window}
\begin{figure}
        \centering
	\includegraphics[width=0.5\textwidth ]{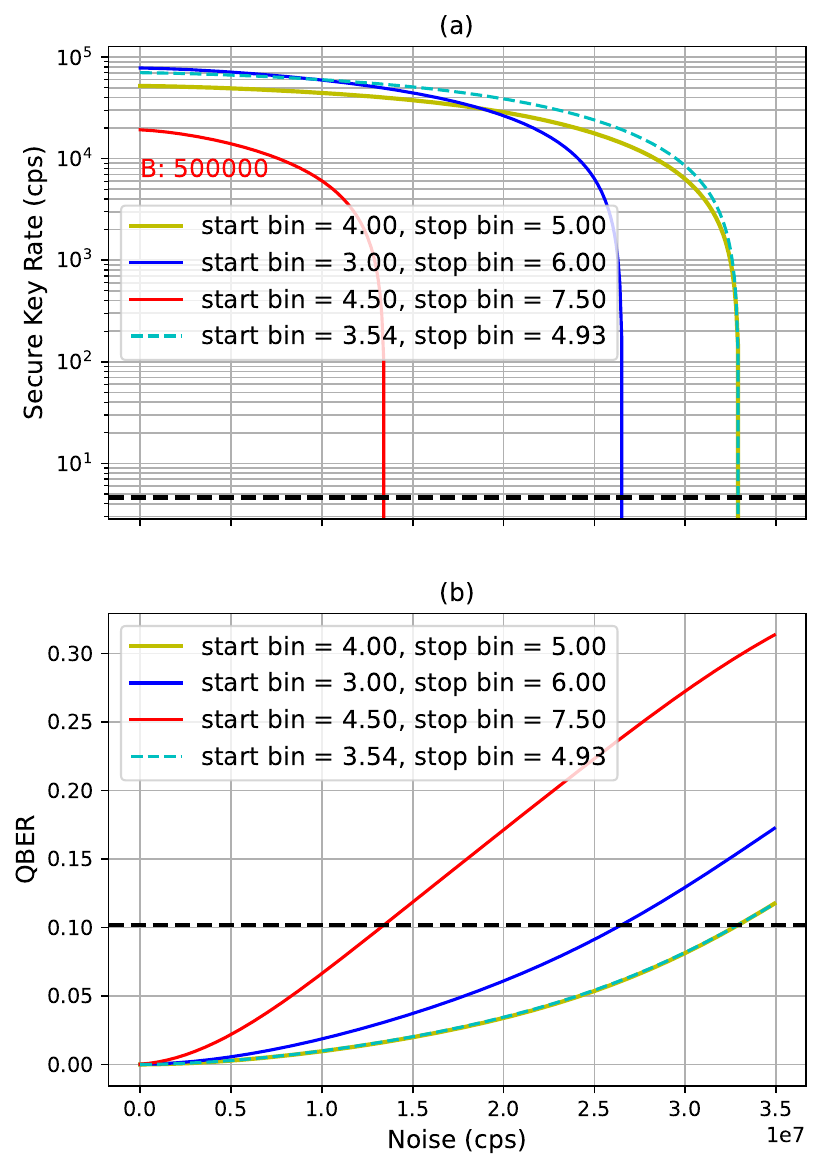}
	\caption{Evaluated QKD parameters of secure key rate and QBER in noisy settings for a set of start and stop bin values. Different start and stop bin choices set the threshold of noise (dark dotted line) for SKR (a) and QBER. The dotted cyan line shows an optimized choice of coincidence window.}
        \label{fig:fig5}
\end{figure}
The coincidence window plays a critical role in determining the secure key rate (SKR) and quantum bit error rate (QBER), particularly in noisy operating conditions. While a narrower coincidence window reduces accidental coincidences and improves QBER at high noise levels, it also discards genuine coincidence events and degrades SKR in low-noise conditions. Therefore, an optimal choice of the coincidence window must balance these competing effects.

To identify the optimal coincidence window for a given photon pair rate and noise level, we numerically optimized the secure key rate $R$ with respect to the start and stop times $(a,b)$ of the coincidence window, using the theoretical model developed in Sec.~\ref{Theory}. The optimization was performed by maximizing $R(a,b)$ under fixed experimental parameters, including the photon pair rate and detector noise levels.

Specifically, we employed a gradient-based optimization algorithm (Broyden–Fletcher–Goldfarb–Shanno, BFGS) to determine the coincidence window that maximizes the SKR for a given noise condition. The optimization procedure treats the coincidence window boundaries $(a,b)$ as free variables while all detector and source parameters are held fixed.

As a representative example, for a photon pair rate of $B=500$~kcps and a total noise level of $22$~Mcps distributed equally among the detectors, the optimization yields an optimal coincidence window of $a = 3.54$~ns and $b = 4.93$~ns. With this choice, the system achieves a maximum secure key rate of 33,162 bits and tolerates noise levels up to $32$~Mcps before the SKR drops to zero. The corresponding trends in SKR and QBER for different coincidence window choices are shown in Fig.~\ref{fig:fig5}.
\section{Conclusion}\label{conclution}
In this paper, we propose an optimization method that allows efficient execution of entanglement-based QKD protocols in a noisy environment. We begin by considering a single-basis experimental setup to understand the factors affecting key QKD parameters in a CW-driven system. Although a single-basis system is insufficient for executing a complete QKD protocol, it simplifies the learning of individual factors that affect the system. We specifically investigate the factors influencing the nonideal behavior of detectors in continuous-wave (CW) driven QKD, identifying elements affecting practical models.

We also presented a test to characterize detector functional changes owing to high photon rates. An experimental test to understand detector functional changes in a pulsed laser-driven QKD has previously been shown \cite{stipvcevic2017advanced}, where changes in detector parameters such as dead time, variation in jitter, and twilight effect were investigated. Contrary to CW-QKD, pulsed lasers offer full control of the pulse time interval, including the adjustment of undesirable detection timing shifts. This would not be straightforward in a CW-driven QKD, particularly because this effect can adversely affect performance if detection rates are imbalanced. In this scenario, our model will be helpful for analysis and optimization.

Incorporating these findings from the single-basis system, we develop a theoretical model applicable to a two-basis system. Given the novelty of this effect, developing an optimization model is crucial. We introduce a simple and efficient model useful for identifying and addressing detector non-idealities that degrade QBER and SKR. The advantage of our developed model is its wide applicability across various QKD protocols, as it is based on the coincidence measurement used in many, if not all, discrete variable entanglement-based QKD systems.

Furthermore, our demonstrated noise-induced shift in detection timing events possibly threatens the overall security of QKD, as it is well known that shifts in coincidence detection event times can be exploited by eavesdroppers for the hacking of secret data \cite{lamas2007breaking}. More studies are needed to understand the extent to which this shift influences the security of the system. One possible solution to mitigate the rate-induced timing shift is to characterize detector efficiency versus input photon rate analysis, as shown in \cite{kuniyil2022noise}, and then discard all communication events when the photon rate exceeds the onset value of efficiency degradation. From our analysis, we observed the onset of efficiency degradation and coincidence histogram shift after $\sim1$ Mcps (although the rated maximum count rate of the detector used is much higher, $\sim38.1$ Mcps). $1 Mcps$ is the value manufacturer tested the dead time. From our method we found that default dead time of the detector is $\sim 23$ ns. this is the exact dead time provided by the manufacturer.

This study enables researchers to implement efficient strategies prior to system deployment. Furthermore, our findings will be valuable for research in quantum detection technologies. Finally, our analysis of the detector's jitter and its impact on photon detections in the time domain is essential for the development and assessment of time-bin-encoded QKD protocols \cite{jin2019genuine}.
\appendix
\section{Delay-Scan Coincidence Histogram from Time-Tagged Events} \label{app_A}

Coincidence measurements were computed from raw time-tagged detection events using a delay-scan algorithm. The input data consist of a sequence of time tags and detector identifiers stored in a CSV file with the structure:

\begin{center}
\texttt{Time, Detector}
\end{center}

where \texttt{Time} denotes the recorded time tag (integer-valued) and \texttt{Detector} specifies the detector/channel index.

\vspace{0.5em}

The coincidence curve $C(\Delta)$ is obtained by applying a discrete delay $\Delta$ to all timestamps from detector 1 and counting the number of events that coincide with timestamps from detector 2. Formally,

\begin{equation}
C(\Delta) = \sum_{t_i \in \mathcal{T}_1} 
\mathbf{1}\big( t_i + \Delta \in \mathcal{T}_2 \big),
\end{equation}

where $\mathcal{T}_1$ and $\mathcal{T}_2$ denote the sets of timestamps recorded by detector 1 and detector 2, respectively, and $\mathbf{1}(\cdot)$ is the indicator function.

\vspace{0.5em}

\textbf{Important note.}  
This implementation defines coincidences using \emph{exact timestamp equality}. This approach is valid when timestamps are discretized into fixed-width time bins and coincidence is defined by bin alignment. For experiments requiring a finite coincidence window (e.g., $|t_1 + \Delta - t_2| \leq \tau$), the matching condition should be modified accordingly.

\vspace{1em}

The following Python script was used to compute the coincidence histogram:

\begin{lstlisting}[language=Python]
import numpy as np
import pandas as pd
import matplotlib.pyplot as plt

def coincidence_vs_delay(
    csv_path,
    det1=1,
    det2=2,
    delay_start_ps=20,
    delay_stop_ps=80,
    delay_step_ps=4,
):
    """
    Compute coincidence counts C($\Delta$) over a discrete delay scan.
    Coincidence condition: (t_det1 + $\Delta$) == t_det2
    """

    data = pd.read_csv(csv_path)
    data.columns = ["Time", "Detector"]

    t1 = data.loc[data["Detector"] == det1, "Time"].to_numpy()
    t2 = data.loc[data["Detector"] == det2, "Time"].to_numpy()

    t2_set = set(t2)

    delays = np.arange(delay_start_ps, delay_stop_ps, delay_step_ps)

    counts = []
    for d in delays:
        counts.append(sum((t + d) in t2_set for t in t1))

    return delays, np.array(counts)

csv_file = r"F:/07-07-22/20minDataNonoise.csv"

delays, coincidence_counts = coincidence_vs_delay(csv_file)

plt.plot(delays, coincidence_counts, "o-")
plt.xlabel("Delay $\Delta$ (ps)")
plt.ylabel("Coincidence count C($\Delta$)")
plt.grid(True)
plt.show()
\end{lstlisting}

\vspace{0.5em}

This procedure yields the coincidence distribution as a function of relative detector delay, allowing identification of the temporal correlation peak.
\section{Numerical Optimization Procedure}\label{app_B}

The optimization of the coincidence window boundaries was carried out using the Broyden–Fletcher–Goldfarb–Shanno (BFGS) algorithm, a quasi-Newton method for unconstrained optimization. The objective function was defined as the negative of the secure key rate, $f(a,b) = -R(a,b)$, such that minimizing $f$ corresponds to maximizing the SKR.

The optimization variables were the start and stop times $(a,b)$ of the coincidence window. Gradients were evaluated numerically, and the Hessian matrix was updated iteratively following the standard BFGS update rule. The optimization was terminated when the gradient norm fell below a predefined tolerance. This procedure was repeated for different noise levels and photon pair rates.
\bibliography{ref} 
\end{document}